\documentclass[twoside,twocolumn,9pt]{article}
\usepackage{extsizes}
\usepackage[super,sort&compress,comma]{natbib}
\usepackage[version=3]{mhchem}
\usepackage[left=1.5cm, right=1.5cm, top=1.785cm, bottom=2.0cm]{geometry}
\usepackage{balance}
\usepackage{mathptmx}
\usepackage{bm}         
\usepackage{amsmath}
\usepackage{amssymb}
\usepackage{mathtools}
\usepackage{mathrsfs}
\usepackage{siunitx}
\usepackage{sectsty}
\usepackage{graphicx} 
\usepackage{lastpage}
\usepackage[format=plain,justification=justified,singlelinecheck=false,font={stretch=1.125,small,sf},labelfont=bf,labelsep=space]{caption}
\usepackage{float}
\usepackage{fancyhdr}
\usepackage{fnpos}
\usepackage[english]{babel}
\addto{\captionsenglish}{%
  
}
\usepackage{array}
\usepackage{droidsans}
\usepackage{xspace}
\usepackage{charter}
\usepackage[T1]{fontenc}
\usepackage[usenames,dvipsnames]{xcolor}
\usepackage{setspace}
\usepackage[compact]{titlesec}
\usepackage{hyperref}
\usepackage{orcidlink}

\usepackage{epstopdf}

\definecolor{cream}{RGB}{222,217,201}

\def\vec#1{\overrightarrow{#1}}    
\def\etal{~\textit{et al.\xspace}}
\def\LC{LC}
\def\LCs{LCs}
\def\ILC{Iso}
\def\NLC{N}
\def\ChLC{Ch}

\def\SmCsLC{SmC$^\ast$\hspace{-0.1em}}

\def\OnsagerVariational{Onsager's variational principle\xspace}

\def\rep#1#2{$\text{#1}_\text{#2}$}

\begin{document}

\pagestyle{fancy}
\thispagestyle{plain}
\fancypagestyle{plain}{
\renewcommand{\headrulewidth}{0pt}
}

\makeFNbottom
\makeatletter
\renewcommand\LARGE{\@setfontsize\LARGE{15pt}{17}}
\renewcommand\Large{\@setfontsize\Large{12pt}{14}}
\renewcommand\large{\@setfontsize\large{10pt}{12}}
\renewcommand\footnotesize{\@setfontsize\footnotesize{7pt}{10}}
\makeatother

\renewcommand{\thefootnote}{\fnsymbol{footnote}}
\renewcommand\footnoterule{\vspace*{1pt}%
\color{cream}\hrule width 3.5in height 0.4pt \color{black}\vspace*{5pt}} 
\setcounter{secnumdepth}{5}

\makeatletter 
\renewcommand\@biblabel[1]{#1}            
\renewcommand\@makefntext[1]%
{\noindent\makebox[0pt][r]{\@thefnmark\,}#1}
\makeatother 
\renewcommand{\figurename}{\small{Fig.}~}
\sectionfont{\sffamily\Large}
\subsectionfont{\normalsize}
\subsubsectionfont{\bf}
\setstretch{1.125} 
\setlength{\skip\footins}{0.8cm}
\setlength{\footnotesep}{0.25cm}
\setlength{\jot}{10pt}
\titlespacing*{\section}{0pt}{4pt}{4pt}
\titlespacing*{\subsection}{0pt}{15pt}{1pt}

\fancyfoot{}
\fancyfoot[LO,RE]{\vspace{-7.1pt}\includegraphics[height=9pt]{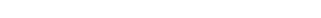}}
\fancyfoot[CO]{\vspace{-7.1pt}\hspace{13.2cm}\includegraphics{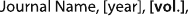}}
\fancyfoot[CE]{\vspace{-7.2pt}\hspace{-14.2cm}\includegraphics{head_foot/RF}}
\fancyfoot[RO]{\footnotesize{\sffamily{1--\pageref{LastPage} ~\textbar  \hspace{2pt}\thepage}}}
\fancyfoot[LE]{\footnotesize{\sffamily{\thepage~\textbar\hspace{3.45cm} 1--\pageref{LastPage}}}}
\fancyhead{}
\renewcommand{\headrulewidth}{0pt} 
\renewcommand{\footrulewidth}{0pt}
\setlength{\arrayrulewidth}{1pt}
\setlength{\columnsep}{6.5mm}
\setlength\bibsep{1pt}

\makeatletter 
\newlength{\figrulesep} 
\setlength{\figrulesep}{0.5\textfloatsep} 

\newcommand{\topfigrule}{\vspace*{-1pt}%
\noindent{\color{cream}\rule[-\figrulesep]{\columnwidth}{1.5pt}} }

\newcommand{\botfigrule}{\vspace*{-2pt}%
\noindent{\color{cream}\rule[\figrulesep]{\columnwidth}{1.5pt}} }

\newcommand{\dblfigrule}{\vspace*{-1pt}%
\noindent{\color{cream}\rule[-\figrulesep]{\textwidth}{1.5pt}} }

\makeatother

\newcommand{\FacultyWaseda}{
    Faculty of Science and Engineering, Waseda University, 3-4-1 Okubo, Shinjuku, Tokyo, 169-8555, Japan\xspace
}

\newcommand{\GradWaseda}{
    Graduate School of Advanced Science and Engineering, Waseda University, TWIns, 2-2 Wakamatsu-cho, Shinjuku, Tokyo, 162-8480, Japan\xspace
}

\newcommand{\JSPSResearchFellow}{
    Research Fellow of Japan Society for the Promotion of Science, Waseda University, 3-4-1 Okubo, Shinjuku, Tokyo, 169-8555, Japan\xspace
}

\newcommand{\CompResOrg}{
    Comprehensive Research Organization, Waseda University, TWIns, 2-2 Wakamatsu-cho, Shinjuku, Tokyo, 162-8480, Japan\xspace
}

\newcommand{\SKCM}{
    International Institute for Sustainability with Knotted Chiral Meta Matter (WPI-SKCM$^2$),
    Hiroshima University, 1-3-1 Kagamiyama, Higashi-Hiroshima, Hiroshima 739-8526, Japan\xspace
}

\newcommand{\ST}{Shunsuke~Takano\orcidlink{0009-0002-9428-4851}}
\newcommand{\TN}{Takuya~Nakanishi\orcidlink{0000-0002-1172-718X}}
\newcommand{\KN}{Kenta~Nakagawa\orcidlink{0000-0002-0577-537X}}
\newcommand{\TA}{Toru~Asahi\orcidlink{0000-0003-4925-8259}}
\twocolumn[
  \begin{@twocolumnfalse}
{\includegraphics[height=30pt]{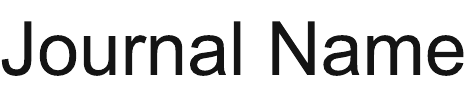}\hfill\raisebox{0pt}[0pt][0pt]{\includegraphics[height=55pt]{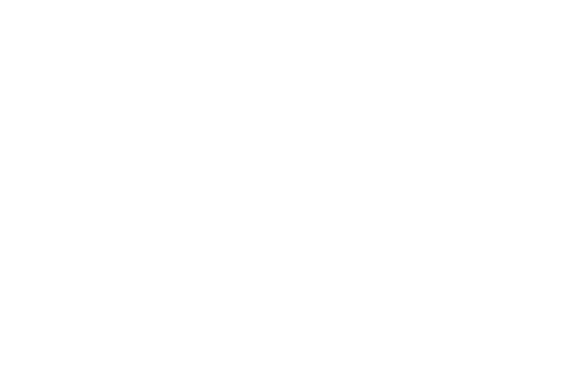}}\\[1ex]
\includegraphics[width=18.5cm]{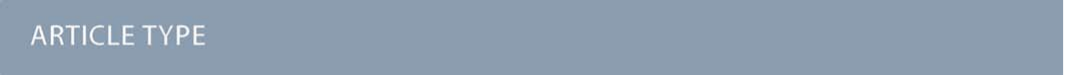}}\par
\vspace{1em}
\sffamily
\begin{tabular}{m{4.5cm} p{13.5cm} }

\includegraphics{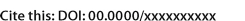} & \LARGE{\textbf{\noindent
    Reversal in thermally driven rotation of chiral liquid crystal droplets
    }} \\
\vspace{0.3cm} & \vspace{0.3cm} \\

 & \large{\noindent
    \ST,\textit{$^{a,b,c}$} 
    \TN,\textit{$^{d}$} 
    \KN\textit{$^{d}$}
    and \TA$^{\ast}$\textit{$^{d,e}$}
    } \\

\includegraphics{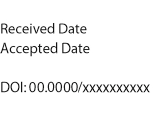} & \normalsize{\noindent
    Thermomechanical coupling in chiral liquid crystals enables the direct conversion of heat current into mechanical rotation, providing a promising mechanism for the utilisation of low-grade thermal energy and heat-driven soft micro-actuation. However, the physical origin governing the direction and magnitude of this coupling remains elusive. Here, we demonstrate that cholesteric liquid crystal droplets undergo a reversal of their rotational direction with changing droplet size and temperature, even under a fixed temperature gradient and unchanged molecular chirality. This previously unrecognised behaviour cannot be accounted for by the conventional description of the thermal Leslie effect. Systematic investigations of droplet size, temperature and molecular structure, together with phenomenological analysis, reveal that thermomechanical coupling is strongly influenced by molecular orientational order and may even reverse sign with changes in the scalar order parameter. These findings identify molecular ordering as an active design parameter, rather than merely a structural descriptor, for controlling thermomechanical energy conversion. More broadly, our results suggest that thermomechanical coupling should be understood as a property emerging from molecular ordering, providing a new framework for designing heat-to-motion energy conversion in chiral soft matter.
    } \\

\end{tabular}

 \end{@twocolumnfalse} \vspace{0.6cm}

  ]

\renewcommand*\rmdefault{bch}\normalfont\upshape
\rmfamily
\section*{}
\vspace{-1cm}


\footnotetext{\textit{$^{a}$~\SKCM.}}
\footnotetext{\textit{$^{b}$~\GradWaseda.}}
\footnotetext{\textit{$^{c}$~\JSPSResearchFellow.}}
\footnotetext{\textit{$^{d}$~\CompResOrg.}}
\footnotetext{\textit{$^{e}$~\FacultyWaseda. E-mail: tasahi@waseda.jp}}

\footnotetext{\dag~Supplementary Information available: the experimental and computational methodology, sample composition, modelling derivation, and other supporting information for this article. See also Zenodo repository\cite{seeZenodo} with DOI:~10.5281/zenodo.22112415.}



\section{Introduction}

    Energy harvesting has been pursued through a variety of principles, and new devices, including thermoelectric devices\cite{Kondo2024NatCommun, Hosseini2026JMaterChemA}, photovoltaic cells\cite{Kojima2009JAmChemSoc}, and quantum-dot heat engines\cite{Josefsson2018NatureNanotech, Jaliel2019PhysRevLett}, have been developed. While these approaches have demonstrated remarkable progress, they are often constrained either by the requirement for large temperature differences, the use of toxic or scarce elements, or the necessity of cryogenic environments to ensure stable operation. These limitations motivate the exploration of alternative platforms that autonomously convert small temperature gradients into mechanical or electrical work under realistic ambient conditions.
    
    A chiral liquid crystal (\LC) is an attractive class of soft matter that offers such a possibility. \LCs~are uniform phases that typically emerging over a temperature range intermediate between crystalline solids and isotropic liquids. The anisotropic molecules retain fluidity while exhibiting long-range orientational order\cite{de_Gennes}. Their orientational order is described by the director vector $\vec{n}$, representing the average molecular orientation, and by the scalar order parameter $S$, which quantifies the degree of alignment. When molecular chirality is present---either intrinsically or through chiral dopants added to an achiral host nematic (\NLC) \LC---the inversion symmetry of the phase is broken, enabling cross-coupled non-equilibrium responses that are forbidden in achiral systems.
    
    One striking consequence of this broken symmetry is Lehmann rotation, a continuous rotation of chiral \LCs\ driven by diffusive or heat currents\cite{Lehmann1900, de_Gennes, Oswald2008PRL, Yoshioka2018}. Since its experimental confirmation, Lehmann rotation has been studied as a model molecular propeller and as a candidate mechanism for active soft devices and micromotors\cite{Tabe2003, Sarman2016, Ignes-Mullol2016prl, Bono2018, Bono2019, Sakai2020, Zhang2021NatRevMater, Oswald2022PhysRevE}. Importantly, Lehmann rotation is autonomous: no cyclic external driving is required, and a steady temperature gradient alone suffices to sustain continuous mechanical motion. From the perspective of energy conversion, this autonomy distinguishes chiral \LCs\ from many microscale heat engines, which often rely on externally imposed, time-dependent protocols.
    
    Microscale heat engines are known to efficiently convert small temperature differences into mechanical work and, in rare events, may even transiently exceed the Carnot efficiency\cite{Martinez2016NaturePhys}. However, as system size decreases, thermal fluctuations increasingly obscure extraction of useful work\cite{Blickle2012NaturePhys}. Quantum and mesoscopic heat engines therefore typically operate only under cryogenic conditions, where fluctuations are sufficiently suppressed\cite{Josefsson2018NatureNanotech}. In contrast, \LCs\ exhibit strong molecular correlations that persist at room temperature, allowing collective molecular motion to manifest as macroscopic rotation that is robust against thermal noise\cite{Tabe2003}. Indeed, Lehmann rotation has been observed under temperature differences as small as $10^2\ \unit{\milli\kelvin}$\cite{Takano2023}, underscoring the suitability of chiral \LCs\ for harvesting low-grade waste heat under ambient conditions.
    
    Theoretical descriptions of Lehmann rotation are commonly formulated in terms of the thermal Leslie effect, a cross-correlation response\cite{Leslie1968II, de_Gennes} in which a true vector, such as a heat current, induces a pseudovectorial torque through chiral symmetry breaking. Within this framework, the torque is expected to be proportional to the applied temperature gradient and to material-specific thermomechanical coefficients. While this phenomenology successfully captures the existence of thermally induced rotation, its manifestation depends sensitively on boundary conditions. In systems where hydrodynamic flow is suppressed, such as slab geometries, the Leslie effect predominantly induces orientational (director) rotation\cite{Oswald2008PhysRevEsliding}. By contrast, when mechanical constraints are relaxed, the same torque may give rise to rigid-body rotation of the \LC\ as a whole\cite{Yoshioka2014, Ito2016, Bono2018, Nishiyama2019, Bono2020, Nishiyama2021SoftMatter, Takano2023}. 
    Another unresolved issue concerns the control of rotational behaviour. Traditionally, the rotation direction of Lehmann rotation has been considered to be fixed by the handedness of the chiral dopant or by the sign of the applied thermal gradient. This view implicitly assumes that the Leslie torque depends only on the director field $\vec{n}$ and on material chirality, while neglecting possible dependencies on other order parameters intrinsic to the \LC\ phase. Given that the Leslie effect is specific to the nematic order, it is natural to expect that its coupling coefficients may also depend on the scalar order parameter $S$\cite{Takano2025PhysRevE}. However, experimental evidence for such a dependence has remained elusive.
    
    In the present work, we investigate steady thermally driven rotation of \LC\ droplets dispersed in glycerol under an applied temperature gradient. By employing a surface-cleaning protocol that prevents droplet adhesion to the glass substrate, we realise a system in which \LC\ droplets remain levitated and are free to undergo rigid-body rotation without violating the non-slip condition. This geometry enables us to directly detect rigid-body rotation induced by the thermal Leslie effect and to quantitatively compare the extracted thermomechanical coefficients with those obtained in slab geometries. Furthermore, by systematically varying the droplet radius and the ambient temperature, we uncover controllable reversals of the rotation direction that cannot be explained by changes in chirality, the direction of the thermal gradient, or the cholesteric (\ChLC) pitch. Our results demonstrate that an extended description of the Leslie effect, incorporating additional torque contributions and a dependence on the scalar order parameter $S$, is required to consistently account for the observed behaviour.

\section{Results and discussion}
\begin{figure}[btp]
\centering
    \includegraphics[width=8.3cm, keepaspectratio]{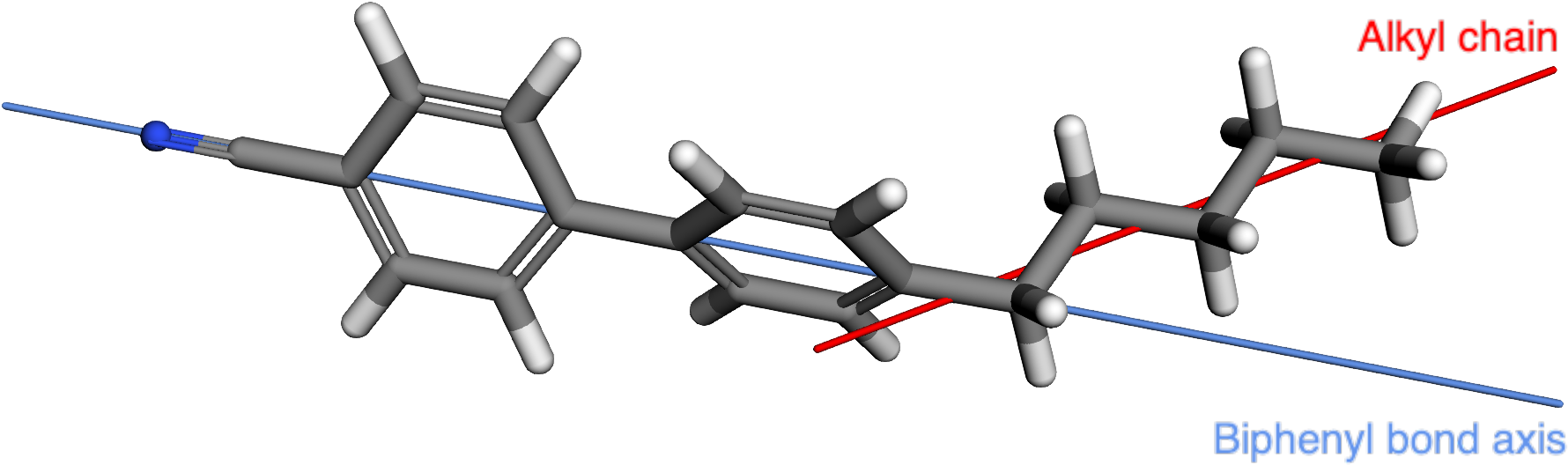}
    \caption{
        Schematic three-dimensional structure of a 5CB molecule optimised at the B3LYP/6-31G(d,p) level. The blue and red lines indicate the biphenyl bond axis and the least-squares regression line fitted to the carbon atoms of the alkyl chain, respectively. Even in the all-trans conformation, the alkyl chain is not collinear with the biphenyl bond axis, forming an angle of ca. $35\ \unit{deg.}$. See also an interactive 3D visualisation via Zenodo\cite{seeZenodo}.
    }
    \label{fig:conf_alkyl}
\end{figure}

\begin{figure*}[tbh]
\centering
    \includegraphics[width=17.1cm, keepaspectratio]{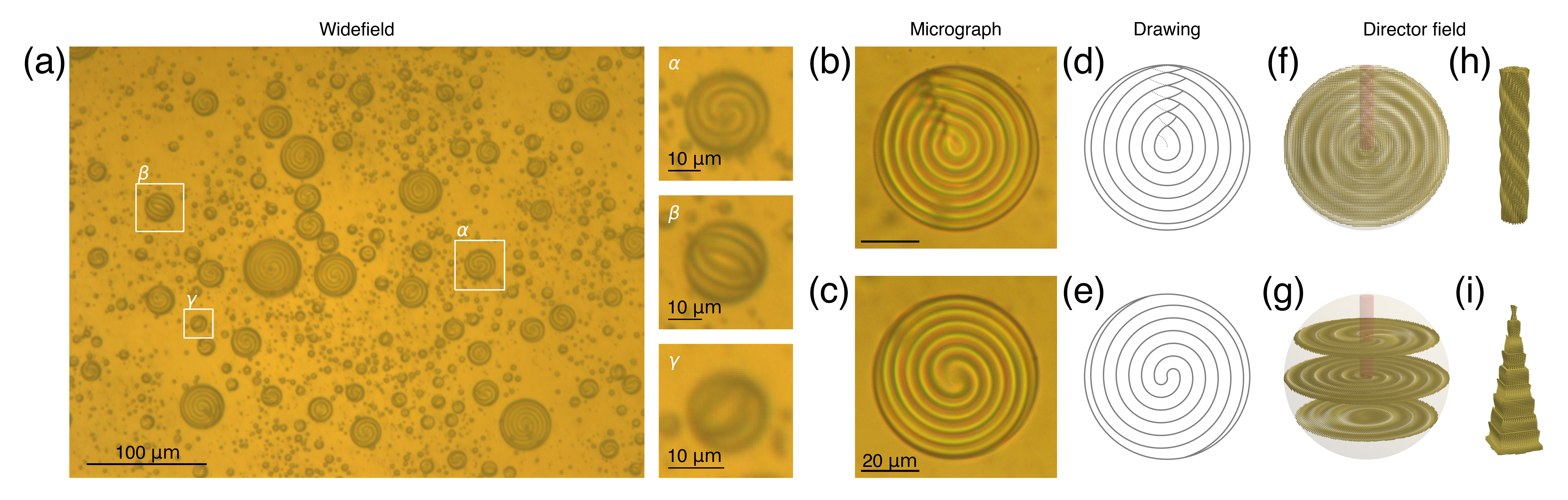}
    \caption{
        (a) Textures of 6CB(\textit{R}) droplets dispersed in glycerol under a temperature gradient. In magnified insets, \ensuremath{\alpha}, \ensuremath{\beta}, and \ensuremath{\gamma} droplets had the radial spherical structure (RSS), the diametrical spherical structure, and the bipolar structure, respectively (See also Movie~S2\cite{seeZenodo}).
        (b) and (c) Micrographs of 7CB(\textit{R}) RSS droplets taken in absence of temperature gradients by detaching the chamber from temperature controllers. 
        (d) and (e) Drawings of the texture when observed from two different orthogonal angles. 
        (f)--(i) Director field under Ansatz that constrains director at droplet surface to tangent plane, derived in Ref.\cite{Bezic1992} (See also Movie~S3\cite{seeZenodo}). Director is visualised as a cylinder. Red region represents the singular $\chi^{+2}$ disclination. 
        (f) Concentric director field near plane where disclination lies. 
        (g) Spiral director field on sections orthogonal to disclination. 
        (h) Cylindrical region around the disclination. 
        (i) Pyramid region opposite disclination across centre of the droplet. Helical director field forms layers with their normal along radial axis.
    }
    \label{fig:geometry_structure}
\end{figure*}

    We observed \LC~droplets dispersed in glycerol. \ChLC\LC\ mixtures $n$CB(\textit{X}) were prepared by mixing a \NLC\LC\ 4$'$-alkyl-4-cyanobiphenyl $n$CB ($n = 5,6,7$), with a chiral dopant (\textit{X})-2-octyl 4-[4-(hexyloxy)benzoyloxy]benzoate ($X = R, S$) at a concentration of $\text{ca.}\, 1.5\%\ \text{w/w}$. The $n$CB homologous series, comprising a rigid biphenyl mesogenic core and a flexible normal alkyl chain~(Fig.~\ref{fig:conf_alkyl}), exhibits the \NLC~phase at or near room temperature~\cite{Gray1975} and has been widely used as a benchmark system in studies of Lehmann rotation~\cite{Eber1982,Oswald2014epl,Oswald2019softmatter,Yoshioka2024}. An immiscible emulsion of glycerol and \ChLC\LCs, in which low miscibility effectively suppresses Marangoni convection, was prepared. The emulsion was confined within a cell made of bare glass substrates with hydrophilic surfaces. See the Supplementary Information (SI)\footnotemark[2] and our previous work\cite{Takano2023} for more technical details. The hydrophilic glass surface prevented the adhesion of LC droplets to the substrate, which was confirmed by the confocal laser scanning microscopy (See also the supplementary Fig.~S3\footnotemark[2] and Movie~S1\cite{seeZenodo}). Glycerol imposes planar degenerate anchoring on the droplets, resulting in three distinct structures (Fig.~\ref{fig:geometry_structure}a\footnotemark[2] and Movie~S2\cite{seeZenodo}). We focused on the radial spherical structure (RSS), which predominates in \LC\ droplets with a radius larger than the cholesteric pitch $p_0$.

    The RSS comprises a spherical bulk formed by the concentric cholesteric layers and a radial disclination\cite{Kurik1982m} (Fig.~\ref{fig:geometry_structure}b--e). The planar anchoring and the similar assumption on bulk orientation derive a qualitative director distribution\cite{Bezic1992}(Fig.~\ref{fig:geometry_structure}f--i and Movie~S3\cite{seeZenodo}). The central spiral texture appears when viewed along the extension of the disclination (Fig.~\ref{fig:geometry_structure}c and e), and the sense of the spiral is designated as either \textsf{S}-shaped or its mirror image\cite{Lehmann1917Tropfen}. The sense was uniquely determined under a fixed temperature gradient $\vec{\nabla} T$ and the handedness of chiral dopant: all RSS droplets of $n$CB(\textit{R}), exhibiting a right-handed cholesteric helix, showed the \textsf{S}-shape under a vertically upward temperature gradient. This spiral sense indicates that the disclination is located toward the hotter side, thereby suppressing the loss of free energy associated with the distorted orientational order at the disclination\footnotemark[2].

    In following experiments, a temperature gradient was applied to the chamber vertically upward (defined as the $z$ axis) by heating the top and cooling the bottom of the chamber, and Rayleigh--B\'{e}nard convection, in which heated fluid floats\cite{Benard1901, Rayleigh1916}, was prevented.
    
\subsection{Hydrodynamic flow inside and outside a droplet}
\subsubsection{Rigid-body rotation detected by PTV.~~}
\begin{figure}[bth]
\centering
    \includegraphics[width=8.3cm, keepaspectratio]{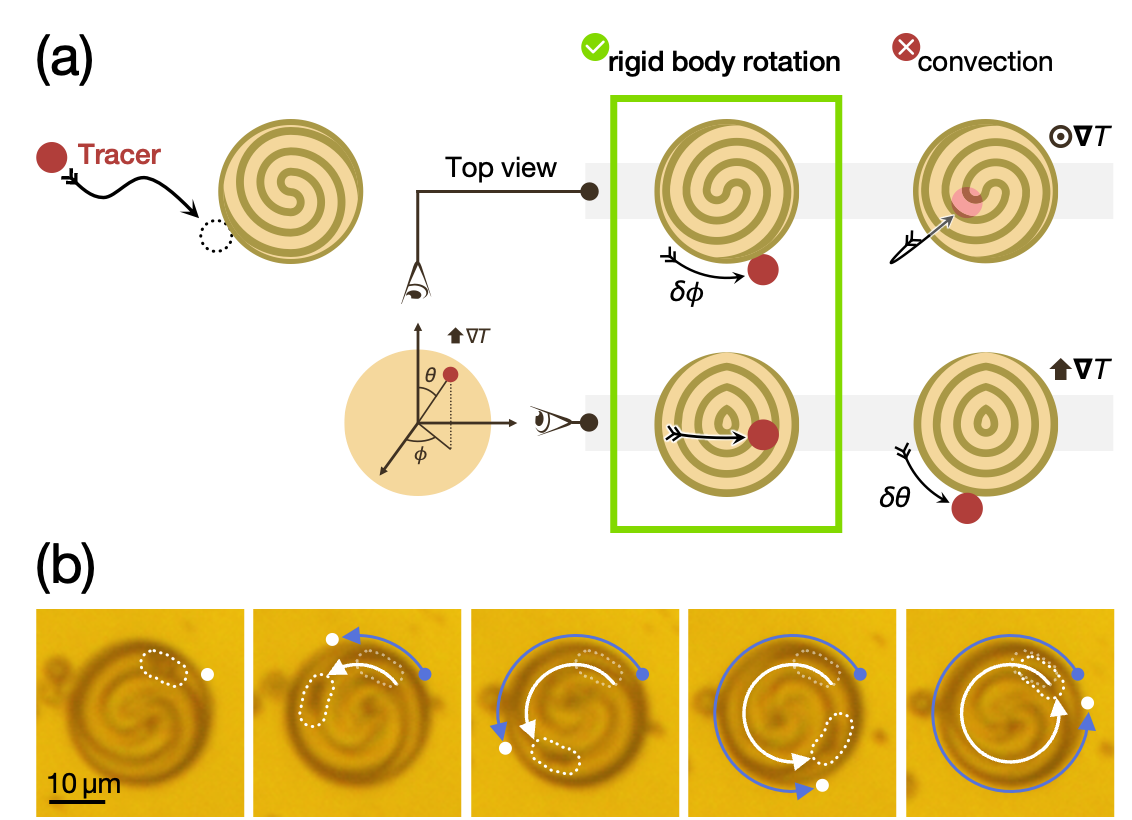}
    \caption{
        Rigid-body rotation identified by PTV. 
        (a) Displacement of a tracer induced by rigid-body rotation and convective flow. The upper panels show a top view, whereas the lower panels show how the motion appears from the side.
        (b) A rotating 7CB(\textit{R}) droplet with tracers on its surface (See also Movie~S4\cite{seeZenodo}). The tracers enclosed by the white dotted line rotated together with the texture. The transparent dotted line designates the position of tracers at the initial state. The orientations of the droplet at each moment and the initial state are represented by a white and blue points, respectively. Images are shown every $2700\ \unit{s}$, and the contrast is enhanced for better visibility.
    }
    \label{fig:rigidbody}
\end{figure}

    The hydrodynamic behaviour of \LC\ droplets was characterised using particle-tracking velocimetry (PTV)\cite{Nishiyama2019} (Fig.~\ref{fig:rigidbody}a). Polystyrene particles (diameter $2.5\ \unit{\micro\metre}$) were used as tracers by adding to an emulsion sample. By stirring the sample, some particles adhered to the droplet surfaces. Owing to the slightly high interfacial tension between glycerol and $n$CB ($14\ \si{\milli\newton\per\metre}$\cite{Oswald2019softmatter}), the particles remained attached to the droplet surface while experiment. At the fluid–solid interface, the no-slip condition applies; thus, the particles are advected at the velocity of the background flow.
    
    We observed \LC\ droplets bearing several adhered tracers (Fig.~\ref{fig:rigidbody}b and Movie~S4\cite{seeZenodo}). The tracers did not reside at the disclination, and no orientational distortions were detected around them, indicating that they are trapped by interfacial energy. The texture of the droplet rotated as a rigid body. The tracers were simultaneously advected in the same direction with identical angular velocity, indicating that the droplet underwent rigid-body rotation. This observation precludes any apparent rotation arising from Marangoni convection or director rotation (Fig.~\ref{fig:rigidbody}a). Since the tracers exhibited no displacement in the $z$ direction, convection was absent. Any motion along $z$ would be projected as movement either away from or toward the droplet centre. Thus, the droplet rotation is driven by cross-couplings rather than by convection. Were the rotation due to director reorientation, it would not involve background flow, and the tracers would remain stationary. The observed tracer behaviour agrees with previous reports\cite{Takano2023}. 
    
    Hydrodynamic flow arises due to antisymmetric stresses acting on the fluidic elements of the LCs\footnote[3]{The authors thank T. Arima for the thoughtful comment.}. The thermal Leslie effect causes heat current to generate antisymmetric (and symmetric) stress and hence constitutes a plausible mechanism for the observed droplet rotation. The molecular field is also induced by Leslie effects, which in turn partially contributes to the antisymmetric stress\cite{Leslie1968II, Takano2025PhysRevE}. That is, director rotation necessarily involves flow, provided that such flow is not suppressed by boundary conditions such as adhesion to the substrate. The droplets in glycerol are freely suspended, without adhesion to the substrate; consequently, flow is permitted. Under such conditions, it is natural for the Leslie effect to induce rigid-body rotation. AZ effects also induce antisymmetric stress. Rigid-body rotation is supported both by experimental observations and by the underlying mechanism. Importantly, the motion of the tracers directly demonstrates that the thermally driven rotation converts thermal energy into work.

\begin{figure*}[btp]
\centering
    \includegraphics[width=17.1cm, keepaspectratio]{graph/flow_panels.png}
    \caption{
        Vorticity $\omega$ obtained from numerical solutions of the Navier--Stokes equation. 
        (a) Steady-state vorticity field around a droplet located near a substrate. Coordinates $r$ and $z$ are normalised by the droplet radius $R$. The white and black dashed curves denote the substrate surface and droplet surface, respectively.
        (b) Colour bar for panels (a), (c), and (d).
        (c) Magnified view of the region in (a) close to the substrate.
        (d) Approximate vorticity field corresponding to (c), modelled as a Couette flow $\omega \simeq \omega_0\,  z/\sqrt{1 + d - \sqrt{1 - r^2}}$.
        (e) Time evolution of the torque $T_d$ computed from the vorticity field $\omega$. The substrate distance $d$ for each curve is indicated in panel (f). The grey region indicates the relaxation process towards the steady stated indicated by the white region.
        (f) Dependence of $T_d$ on $d$. Coloured plots show simulated torque at in the steady state ($t > 3$), and the black line represents the fitting curve $T_d / T_\infty = 1.1942(3) - 0.382(8) d \log{d} + 0.07(2) d$.
        (g) Integrand of $T_d$ as a function of the polar angle $\theta$. Red solid line: numerical result. Lightcoral solid line: approximation near the substrate. Black dashed line: case without a substrate. 
    }
    \label{fig:flow_panels}
\end{figure*}

\subsubsection{Dragged flow estimated by numerical calculations}
    We estimated the flow field formed around the droplet via numerical calculations. Confocal microscopy revealed that the droplet floated near the top of the cell, making point contact with the substrate\footnotemark[2]\cite{Takano2023}. Intuitively, the torque may increase when the droplet approaches the substrate, but it remains unclear whether the torque would diverge if the droplet–substrate distance $d$ approached zero. Confocal microscopy affords a $z$ resolution of $\sim 1\ \unit{\micro\metre}$. The torque in regions beyond this limit, closer to the substrate, was evaluated via numerical calculations.
    
    Navier--Stokes equation was numerically solved with finite element method under low Reynolds number conditions ($\mathrm{Re} \sim 10^{-9}$), corresponding to the Stokes flow regime. We focused on the $z$ component of the vorticity $\omega$, as the other components are independent on it and do not contribute to the rigid-body rotation. An analytical solution for $\omega$ is known in the absence of a substrate. This analytical solution was applied as the initial state of $\omega$. Figure~\ref{fig:flow_panels}a shows that $\omega$ on the side opposite to the substrate remained nearly unchanged compared to the case without a substrate. So, the influence of the substrate is limited to its vicinity. Near the substrate, $\omega$ was distorted and was well approximated as Couette flow (Figs.~\ref{fig:flow_panels}c and d). That is, the local angular velocity at a given point was proportional to $z$. The $\omega$ to $0.1 \omega_0$ at a distance of droplet radius $R$ from the droplet surface, with $\omega_0$ angular velocity of the droplet. This explains that neighbouring droplets remained essentially unmoved because the angular velocity was (effectively) limited to the vicinity of the substrate.
    
    The torque was computed from $\omega$. Since the initial condition of the calculation assumed the absence of the substrate, the $\omega$ near the substrate rapidly increased as time progressed, resulting in an elevated torque (Fig.~\ref{fig:flow_panels}e). The torque in the steady state $T_d$ was expressed as a function of $d$. The fitting curve in Fig.~\ref{fig:flow_panels}f is derived from an asymptotic analysis under the assumptions that (1) the flow field on the side opposite the substrate is approximated by the analytical solution in the absence of the substrate, and (2) the flow near the substrate can be approximated by Couette flow. With a polar angle $\theta$, the integrand for calculating $T_d$ coincides with that in the absence of the substrate for $\theta > \pi / 2$, indicating that the substrate’s influence is negligible in this region (Fig.~\ref{fig:flow_panels}g). For $\theta < \pi / 2$, the integrand is well expressed by the approximate solution (note that this is not a fitting) but rapidly deviates from the approximation for large $d \sim 0.1$. For $d \simeq 0.01$, the approximation remains accurate up to $\theta \simeq \pi / 4$. The $d \log{d}$ term is of lower order than $d$, dominating the behaviour when $d < 0.1$ and ensuring convergence of $T_d$ to a finite value which is $\text{ca.}\, 12\%$ greater than $T_\infty$ (Fig.~\ref{fig:flow_panels}f).

\subsection{Reversal driven by droplet radius}
\begin{figure}[tbhp]
\centering
    \includegraphics[width=8.3cm, keepaspectratio]{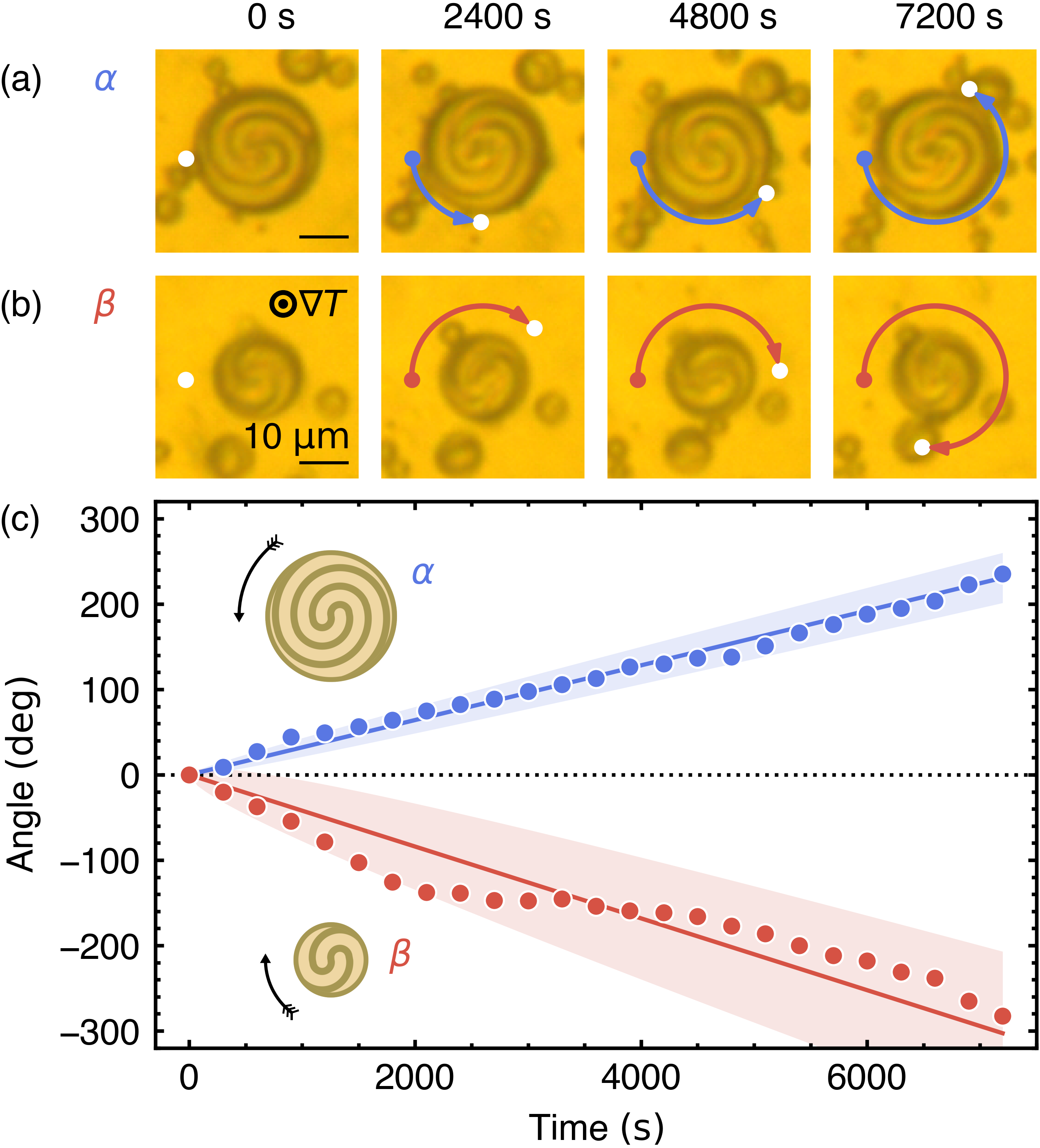}
    \caption{
        Rotation of droplets. 
        (a) and (b) Micrographs of rotating droplets of different sizes, \ensuremath{\alpha} and \ensuremath{\beta}, under temperature gradient $\nabla_z T = +1.00 \times 10^4\ \unit{\kelvin\per\metre}$ (target symbol in (b) is perpendicular to this figure in the direction of the reader). Disclination is oriented parallel to optical axis. Initial and current directions of droplets are represented by coloured and white circles, respectively. See also Movie~S5\cite{seeZenodo}. 
        (c) Rotation angle of droplets \ensuremath{\alpha} (blue) and \ensuremath{\beta} (red) against time. Positive in angle axis represents anticlockwise rotation. Measured rotation angles from initial position are plotted with regression lines. Shaded areas indicate uncertainties of doubled standard deviations due to thermal fluctuations. For smaller droplets, angular variations were more significant due to thermal fluctuations. The contrast of the images was adjusted.
    }
    \label{fig:linear}
\end{figure}

    Figure~\ref{fig:linear}a shows the rotational motion of 7CB(\textit{R}) droplets under a steady temperature gradient (See also Movie~S5\cite{seeZenodo}). As the clean glass substrate prevented the adhesion of hydrophobic LC droplets\cite{Takano2023}, the rotational motion of the droplets was successfully demonstrated (See also Fig.~S3\footnotemark[2] and Movie~S1\cite{seeZenodo}). Remarkably, the rotational direction of the droplets reversed depending on their radii. Figure~\ref{fig:linear}c shows the temporal change in the rotational angle of a large droplet \ensuremath{\alpha} (with radius of $13.0\ \unit{\micro\metre}$) and a small droplet \ensuremath{\beta} (with radius of $9.1\ \unit{\micro\metre}$). Droplets \ensuremath{\alpha} and \ensuremath{\beta} rotated in the opposite directions due to the difference in their radii. As conventionally known, inverting the chirality of sample or the direction of the temperature gradient causes the reversal of rotational direction\cite{Oswald2008PRL, Oswald2021, Oswald2019rev, Takano2023}. However, the present reversal with size (droplet radius) has not been yet predicted and is presented in this article for the first time.

\begin{figure}[bthp]
\centering
    \includegraphics[width=8.3cm, keepaspectratio]{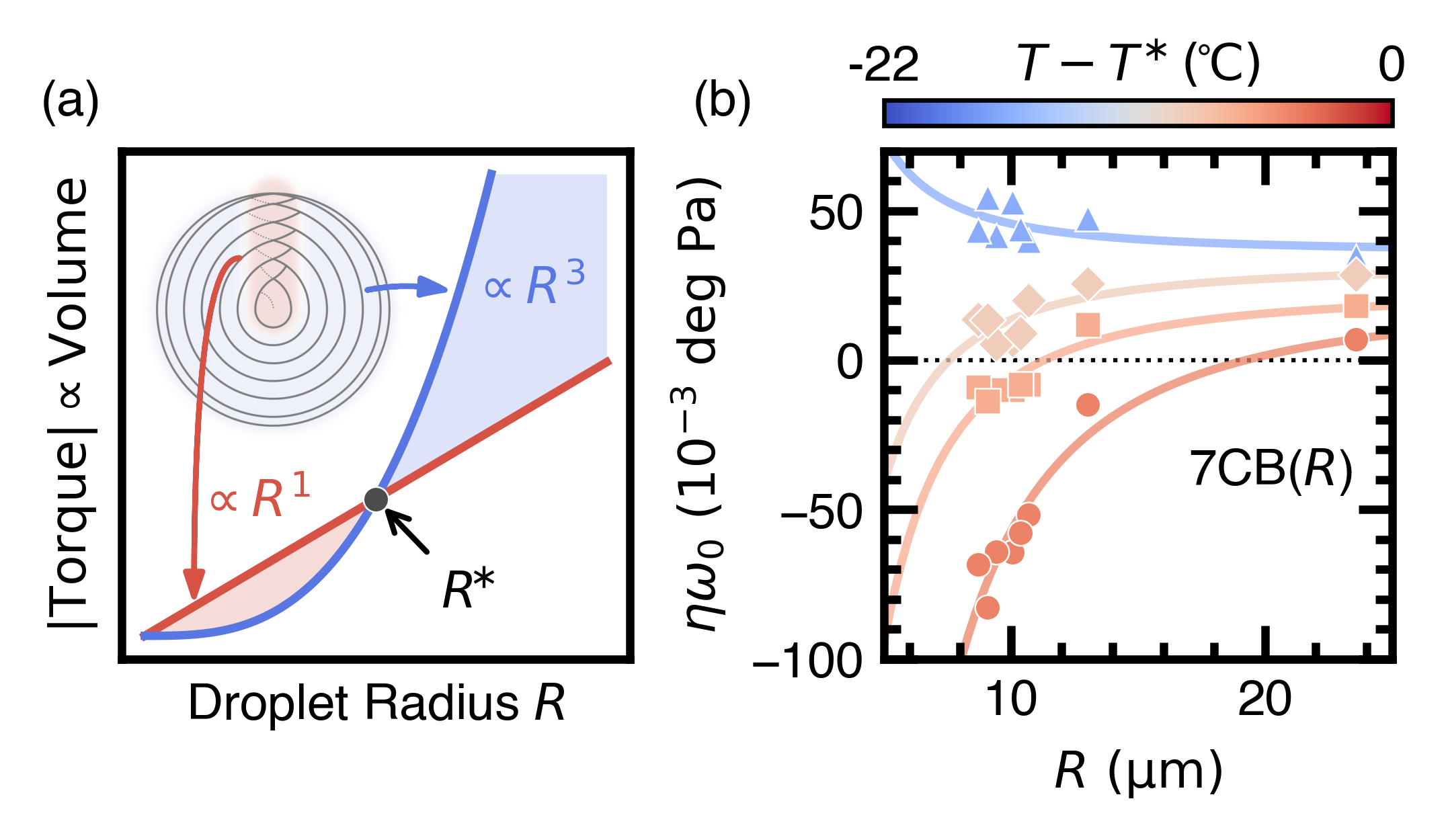}
    \caption{
        (a) Schematic diagram of torque in constituent structures. Blue and red lines represent torque induced in shells (shaded in blue) and disclination (shaded in red), respectively. Each torque is proportional to volume and opposite to each other. Rotational direction of droplets shifts at switching radius $R^\ast$ where torques are in equilibrium. 
        (b) Angular velocity against radius under the temperature gradient $\nabla_z T = + 1.00 \times 10^4\ \unit{\kelvin.\metre^{-1}}$. Angular velocity $\omega_0$ is scaled by viscosity of glycerol $\eta$. Plots are measured values and curves are regression lines obeying $\eta \omega_0 = a_3 - a_1 R^{-2}$, complementing Eq.~\eqref{eq:eta_omega_0} with temperature-dependent fitting parameters $a_3$ and $a_1$ (See also Table~S10 for obtained fitting parameters). Colour of each plot and curve indicates the relative temperature $T - T^\ast$ and corresponds to colour bar above.
    }
    \label{fig:wellknown}
\end{figure}
	
    The reversal of the rotational direction is attributed to the reversal of the torque exerted on the droplet. Two structures that comprise RSS, the spherical bulk and the radial disclination (Fig.~\ref{fig:geometry_structure}b--i), possess different dimensions in terms of extensive property.
    Figure~\ref{fig:wellknown}a schematically depicts the dimension of these structures. The spherical bulk and the radial disclination are three- and quasi-one-dimensional with the volume increasing in proportion to the cubic radius $R^3$ and to the radius $R^1$, respectively. We hypothesised that the Leslie torques occur in opposite directions between them; torques in $n$CB(\textit{R}) are induced anticlockwise and clockwise in the bulk and the disclination, respectively. The torques proportional to each volume compete with each other inside the droplet, and the balance of the competing torques determines the rotational direction of the droplet. A droplet with a radius large enough, in which the bulk is volumetrically dominant, experiences a net anticlockwise torque, whereas a droplet whose radius is so small that the disclination transcends the bulk experiences a net clockwise torque. This description requires that thermal energy is converted into torque differently depending on the microscopic orientational order of the molecules in each structure, which is modelled in Sections~\ref{sec:model_a}--\ref{sec:model_c}.
    
	The macroscopic competition of torque between these inner structures was formulated by the dimensional analysis. Full details of the derivation are available in the SI\footnotemark[2]. A minimal expression of entropy production $\dot{S}_{\text{ent}}$ up to numerical factors is a quadratic form of the variables of angular velocity $\omega_0$ and temperature gradient $\nabla_z T$, as 
    \begin{equation}
    \label{eq:TSdot}
        T\dot{S}_{\text{ent}} = - 2\nu R^3 - {p_0}^2 R^1 \omega_0 \nabla_z T + \eta R^3 {\omega_0}^2 + \kappa T^{-1} R^3 \left(- T^{-1} \nabla_z T\right)^2
        ,
    \end{equation}
    where $\kappa$ is the thermal conductivity. The first term represents the energy converted per unit time through the thermal Leslie effect. The coefficients with dimensions of volume $R^3$ and ${p_0}^2 R^1$ describe the contributions of the three-dimensional spherical bulk and the quasi-one-dimensional disclination, respectively. Adopting \OnsagerVariational\cite{Doi2011}, we obtained $\omega_0$ in the steady state,
    \begin{equation}
    \label{eq:eta_omega_0}
        \eta \omega_0 = \nu \nabla_z T \left(1- \left(R / p_0\right)^{-2}\right)
        ,
    \end{equation}
    which minimises $T\dot{S}_{\text{ent}}$ in Eq.~\eqref{eq:TSdot} under a fixed temperature gradient $\nabla_z T$. The contribution of the disclination, represented by the factor $R^{-2}$, is responsible for the radial dependence and reversal of the droplets' rotation. Eq.~\eqref{eq:eta_omega_0} derives a switching radius $R^\ast$ that is almost equal to the pitch $p_0$. As shown in Fig.~\ref{fig:wellknown}b, Equation~\eqref{eq:eta_omega_0} describes the dependence of the angular velocity $\omega_0$ on the radius $R$ with the reversal of the rotational direction well.

    In the dimensional analysis, we assumed that the molecular order leads to a difference in the direction of the induced torque (Fig.~\ref{fig:wellknown}a), which is natural because the Leslie effects would relate to certain intermolecular correlations as an analogy to viscosity\cite{de_Gennes}. The intermolecular correlation is represented by the director $\vec{n}$ and the scalar OP $S$, where the latter is the magnitude of molecular order and reaches zero in the isotropic (\ILC) phase. The $\vec{n}$-dependence of the Leslie effects have been recognised\cite{Leslie1968II}, and the $S$-dependence and the $\vec{n}$-independent effect have also been indicated recently\cite{Takano2025PhysRevE}. The contribution of $\vec{n}$ and $S$ to Leslie effects will be considered for describing the competing torques in the bulk and the disclination in Sec.~\ref{sec:model_c}. 
    

\subsection{Reversal driven by temperature}\label{sec:reversal_driven_by_temperature}
\begin{figure}[tbhp]
\centering
    \includegraphics[width=8.3cm, keepaspectratio]{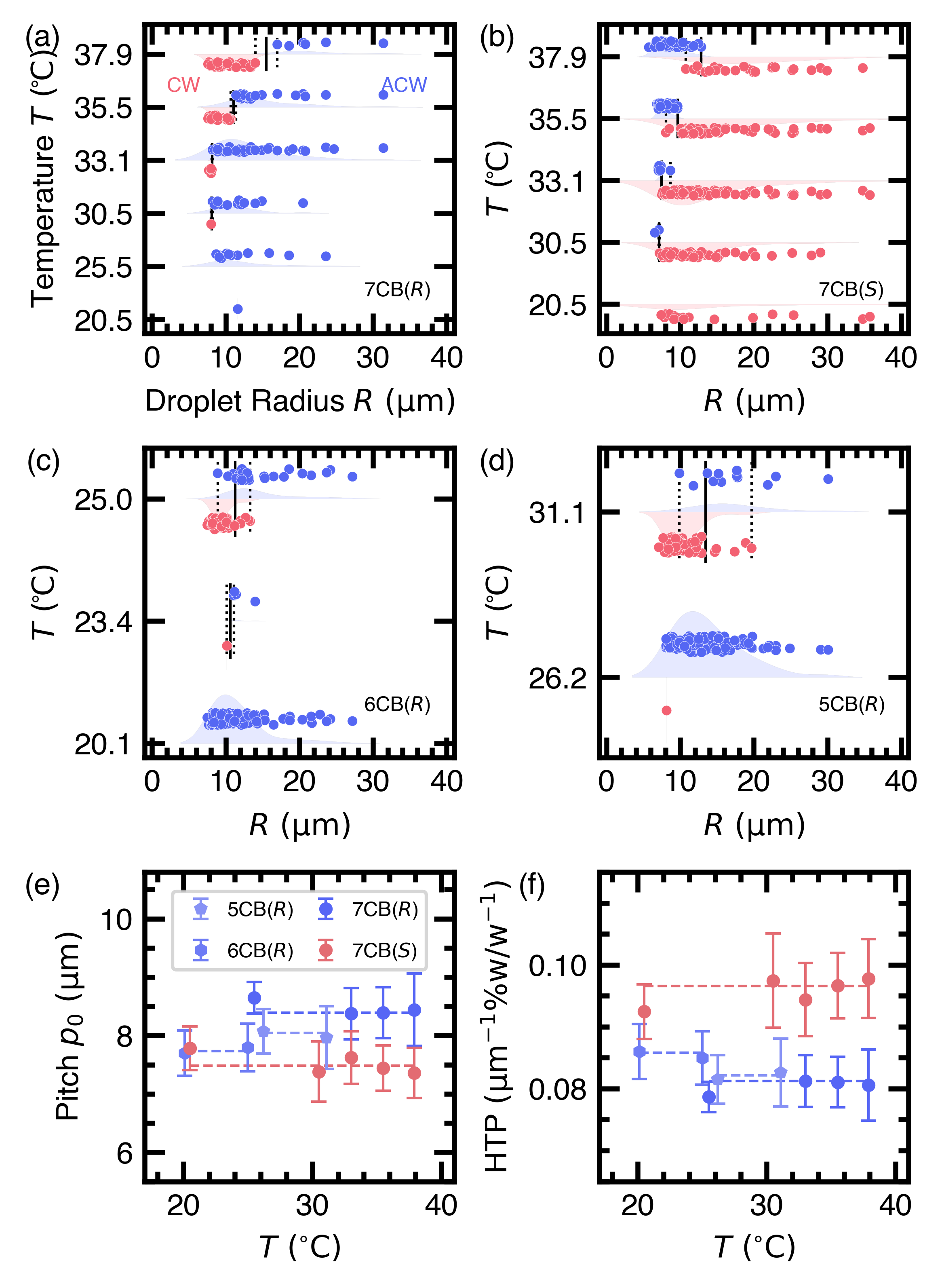}
    \caption{
        (a)--(d) Temperature dependence of rotational direction of droplets of 7CB(\textit{R}), 7CB(\textit{S}), 6CB(\textit{R}), and 5CB(\textit{R}). Anticlockwise (ACW) and clockwise (CW) rotations are plotted in blue and red, respectively. Shaded objects indicate violin plots, showing the frequency distribution of radius across rotational direction and temperature determined by kernel density estimation. Temperatures are displayed as ordered categories rather than on a linear scale. Black solid and dotted lines show the switching radius $R^\ast$ and the range of overlap and/or gap, respectively. 
        (e) and (f) Cholesteric pitch $p_0$ and HTP against temperature, respectively. Error bars represent the $95\%$ confidence interval plus the resolution of the CCD camera (half a pixel).
    }
    \label{fig:sixpanels}
\end{figure}

    As temperature rises, $S$ declines due to thermal fluctuation, and $\nu$ changes accordingly\cite{Takano2025PhysRevE}; thus, the reversal of the rotational direction driven by temperature is also expected. Figures~\ref{fig:sixpanels}a--d show the rotational direction of droplets at different temperatures, clearly demonstrating the above-mentioned reversal. For instance, droplets of 7CB(\textit{R}) with a radius of $\text{ca.}\, 10\ \unit{\micro\metre}$ rotated anticlockwise at $33.1\ \unit{\degreeCelsius}$ but clockwise at temperatures above $35.5\ \unit{\degreeCelsius}$ (Fig.~\ref{fig:sixpanels}a). 
    In the vicinity of $R^\ast$, the droplets rotated sufficiently slowly for the sense of rotation to be perturbed by thermal fluctuations. We therefore defined $R^\ast$ as the (median) radius at which the accuracy\footnote[4]{
        Accuracy is a standard index for binary classification, defined as the fraction of data points correctly classified as exhibiting clockwise or anticlockwise rotation for a given threshold $R^\ast$. The application of Youden’s index yields a closely comparable value of $R^\ast$, indicating that the specific choice of index is not consequential.
    } attains its maximum.
    The switching radius $R^\ast \simeq 10\ \unit{\micro\metre}$ coincides the order of $p_0$, which is consistent with the dimensional analysis in Eq.~\eqref{eq:eta_omega_0}. The $T$-dependent shift of $R^\ast$ indicates that the cross-coupling constant $\nu$ tends to be negative due to the decrease in $S$ at high temperatures and, in contrast, at a sufficiently low temperature, all droplets rotate anticlockwise since $\nu$ is positive due to the higher molecular order even in the disclination. The spiral texture (namely, pitch $p_0$ and handedness of \textsf{S}) was identical at both high and low temperatures. This is consistent with the independence of the pitch and the Leslie effects\cite{Oswald2012epje}. 

    This $T$-dependent reversal was also confirmed in 7CB(\textit{S}). Droplets with a radius of $\text{ca.}\, 10\ \unit{\micro\metre}$ rotated clockwise and anticlockwise at $33.1\ \unit{\degreeCelsius}$ and above $35.5\ \unit{\degreeCelsius}$, respectively (Fig.~\ref{fig:sixpanels}b and Movie~S6\cite{seeZenodo}). The rotational direction at a given temperature and radius was opposite for 7CB(\textit{R}) and 7CB(\textit{S}). Similar reversal of the rotational direction was observed for all the alkyl chain lengths considered, including $n= 5$ and $6$ (Figs.~\ref{fig:sixpanels}c~and~d, and Movies~S7~and~S8\cite{seeZenodo}).
    
	One may concern the effect of $p_0$ on $R^\ast$ but Fig.~\ref{fig:sixpanels}e shows that $p_0$ was independent of temperature. Since $n$CB ($n = 5, 6, 7$) do not exhibit smectic phases\footnotemark[2]\cite{Coles1979MolCrystLiqCryst}, $p_0$ remains constant against temperature. Thus, the reversal is not attributed to $T$-dependent variation of $p_0$. We obtained $p_0 \simeq 10^1\ \unit{\micro\metre}$ and found that $p_0$ was slightly different among all samples. This difference comes from the concentration of chiral dopants~$C$, as shown by the close helical twisting power (HTP $\coloneqq (p_0 C)^{-1}$) of $n$CB(\textit{R}). HTP of 7CB(\textit{S}) was $\text{ca.}\, 10\%$ greater than that of 7CB(\textit{R}), which would be attributed to imperfect enantiopurity of chiral dopant
    \footnote[5]{
        According to the specifications provided by TCI (accessed on 2 May 2024), the specific rotation of (\textit{X})-2-octyl 4-[4-(hexyloxy)benzoyloxy]benzoate ($X = R,S$) was reported as $[\alpha]_{\text{D}}^{20} = \mp 27(2)\ \mathrm{deg}$ ($C = 5$, THF), with the negative and positive values corresponding to $X = R$ and $X = S$, respectively. The reported variations of the specific rotation is sufficiently large to account for a $10\%$ difference in the eantiomeric excess and thus HTP.
    }. 
    
\begin{figure*}[hbtp]
\centering
    \includegraphics[width=17.1cm, keepaspectratio]{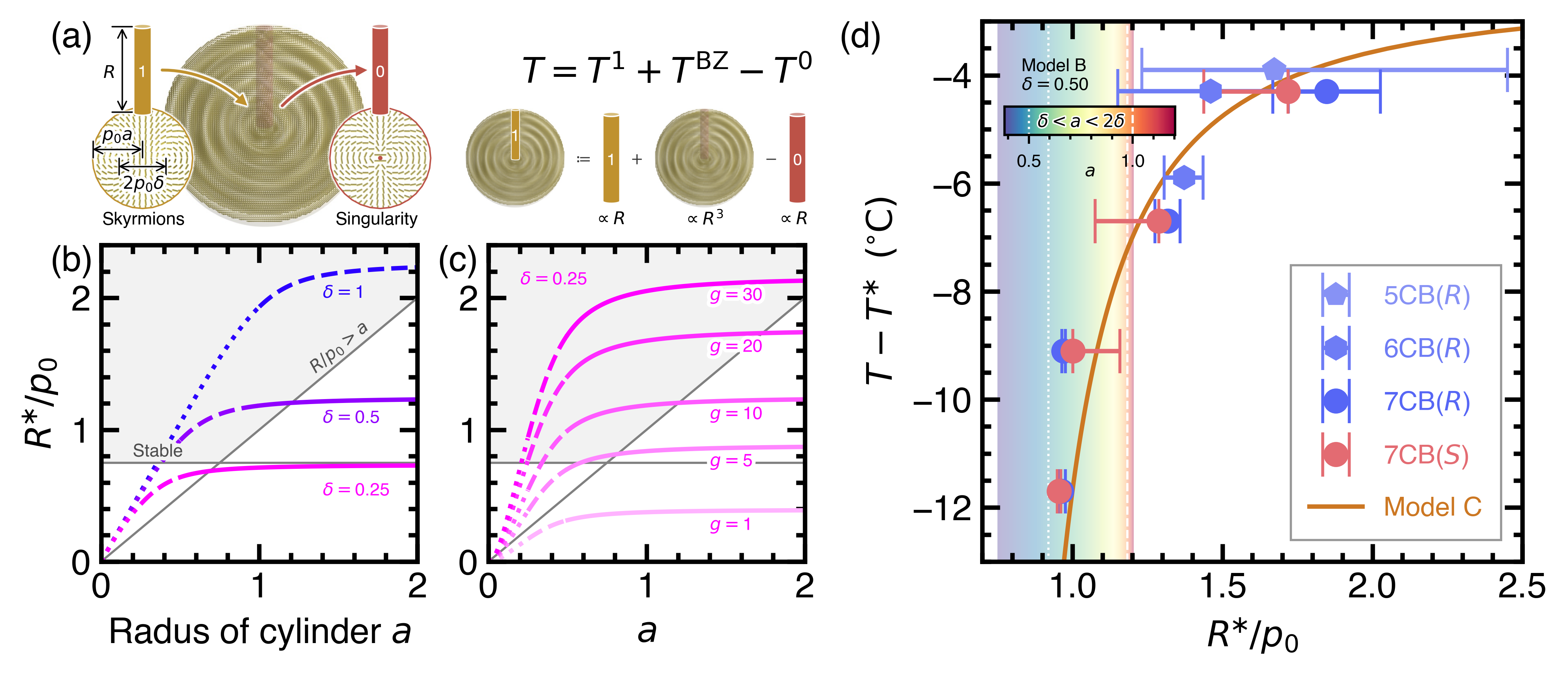}
    \caption{
        Estimation of the switching radius $R^\ast$ considering the thermal AZ and Leslie effects.
        (a) Assumed director field of RSS. The disclination and its vicinity were replaced by a skyrmion pair. The formula describes the computation of net driving torque $T$ exerting on the entire droplet by adding ans subtracting torque of each parts.
        (b) Switching radius $R^\ast$ predicted by Model~B of the AZ effect. The RSS is stable for sufficiently large radius ($R / p_0 \ge 3 / 4$), and the radius of the replaced cylinder never exceeds that of the droplet ($R / p_0 > a$). This feasibility condition is satisfied in the grey region, which indicates the allowable range of $R^\ast$. The dotted line corresponds to $0 \le a \le d$; the dashed line to $d \le a \le 2d$; and the solid line to $2 d \le a$.
        (c) Switching radius obtained with Model~C, which incorporates the $S$-dependence and both the $\vec{n}$-dependent and -independent Leslie effects.
        (d) Plot of $R^\ast$ for each $n$CB(\textit{X}). The brown line represents Model~C in Eq.~\eqref{eq:model_c} with the optimised fitting parameters $S\nu_3 / \nu_2 = -2.3(3) (T^\ast - T + \delta T)^\beta,\ \nu_3' / \nu_3 = -0.45(2)$ as well as the proper hyper-parameters $\delta = 0.25$\cite{Sec2012}, $a = 2\delta = 0.5$, $\delta T = 1\ \unit{\degreeCelsius}$, and $\beta = 0.2$. The coloured band represents the prediction range of Model~B ($\delta = 0.5$). The white dotted and dashed lines represent $a = \delta$ and $a = 2 \delta$, respectively.
    }
    \label{fig:long}
\end{figure*}

    Figure~\ref{fig:long}d shows the switching radius against the temperature. The switching radius $R^\ast$ is normalised by divided by the pitch $p_0$. Temperature $T$ was subtracted by the \ChLC--\ILC~transition temperature $T^\ast$. A common dependence was found among $n$ on $R^\ast / p_0$ and $T - T^\ast$. The universality manifested in $T - T^\ast$ attests that the reversal is governed by the scalar OP $S$ rather than the temperature $T$ itself. Though the mechanism underlying the reversal due to $S$ remains unresolved in the present experiment, the sign reversal of the Leslie cross-coupling coefficients is allowed by symmetry and thermodynamics\cite{Takano2025PhysRevE}. Since thermal energy transport in low-molecular systems is likely governed by rotational degrees of freedom\cite{Ohara1999JChemPhys}, chiral molecules may exhibit a preferential sense of rotation associated with translational heat transport. The rotational motion of the molecules is possibly be influenced by the mobility of the alkyl chains, since in the all-trans conformation the alkyl chains extend away from the biphenyl bond axis (Fig.~\ref{fig:conf_alkyl}) and thus can contribute to rotational degrees of freedom. In the \SmCsLC~phase, conformational transitions induce reversal of the spontaneous polarisation without inversion of the helical pitch\cite{Goodby1987}. Energy exchange between the core and the alkyl chain has been demonstrated in the phonon modes of 5CB\cite{Bolmatov2018JPhysChemLett}. 

\subsection{Model A: original Leslie effect}\label{sec:model_a}
    Model~A considers the director field in Fig.~\ref{fig:long}a and the original Leslie effect. The thermal Leslie effect constitutes a plausible mechanism for thermally driven rotation, yet the original formulation\cite{Leslie1968II} remains insufficient to account for the observed reversal. The Ansatz of the director field bases on the solution by Bezi\'{c}\etal\cite{Bezic1992} $\vec{n}^{\text{BZ}}$, with the disclination revealed by simulation by Se\v{c}\etal\cite{Sec2012}: the disclination is non-singular but rather formed by the intertwining two skyrmion tubes (or a bimeron)\footnotemark[2]. The twist and the intertwining occur in opposite directions\cite{Sec2012}, since points with the same orientation in the cholesteric layer background wind counter to the twist\footnotemark[2]. This relationship is reminiscent of ply yarn. The spatial modulation of $S$ was omitted as the simplest approximation.

    The cylindrical region (with radius $p_0 a$ and height $R$) covering the $\chi^{+2}$ singularity was subtracted by an approximated singular director field $\vec{n}^{0}$, and replaced with intertwined skyrmions with director field $\vec{n}^{1}$ (Fig.~\ref{fig:long}a). The distance between cores of the two skyrmions was $2 p_0 \delta$. Simulation by Se\v{c}\etal~suggests that $\delta \simeq 0.25$, which results in the distance being equal to half pitch $p_0 / 2$. The cores of skyrmions are in the perimeter of the cylinder when $a = \delta$, and the entire skyrimion is inclosed in the cylinder when $a \ge 2\delta$. Thus, most proper choice of parameters is $\delta = 0.25$ and $a = 0.5$.
    
    In the original Leslie torque, its relative direction to $\vec{\nabla} T$ (i.e. the sign of component parallel to $\vec{\nabla} T$) is independent of $\vec{n}$. When $\vec{\nabla} T = \nabla_z T \ \hat{\vec{z}}$, the $z$ component of torque $\tau_z$ is\cite{Leslie1968II,de_Gennes,Takano2025PhysRevE}
    \begin{equation}
        \tau_z
        =
        - [\nu\ \vec{n} \times (\vec{n} \times \nabla_z T\ \hat{\vec{z}})]
        = \nu \left( 1 - (n_z)^2 \right) \nabla_z T.
    \end{equation}
    For any director field, the sense of rotation about $\vec{\nabla} T$ is determined by a single Leslie coefficient $\nu$. The $T$-driven reversal can only be accounted for the $T$ dependence of the cross-coupling coefficient.

    However, Model~A is useful to estimated the Leslie coefficient $\nu$. For simplicity, we assume the singular director field $\vec{n}^{\text{BZ}}$, which is sufficiently good to obtain the order of magnitude of $\nu$. In a sufficiently large droplet, the contribution of the nonsingular disclination is omitted. With the droplet radius $R$, the torque due to Leslie effects is\cite{Takano2023}
    \begin{equation}
        T^{\text{BZ}} = \frac{8}{9}\pi R^3 \nu \nabla_z T
        .
    \end{equation}
    In the steady state, $T^{\text{BZ}}$ and torque due to viscosity $T^{\text{visc.}} = - 8\pi R^3 \omega_0 T_{+0} / T_\infty$ are balanced as $T^{\text{BZ}} + T^{\text{visc.}} = 0$. The hydrodynamic simulation shows $T_{+0} / T_\infty \sim 1.2$ (Fig.~\ref{fig:flow_panels}f).
    Figure~\ref{fig:wellknown}b shows that large droplets at $T - T^\ast = 9.1\ \unit{\degreeCelsius}$ rotate at $\eta \omega_0 \sim + 2\times 10^{-2}\ \unit{deg.\pascal}$. Thus, we estimated $\nu$ as
    \begin{equation}
        \nu
        =
        \frac{9 T_{+0} \eta\omega_0}{T_\infty \nabla_z T}
        \sim
        4 \times 10^{-7}\ \unit{\newton.\metre^{-1}.\kelvin^{-1}}
        .
    \end{equation}
    The obtained $\nu$ is consistent with that measured in previous studies\cite{Oswald2014epl, Oswald2008EPL,  Dequidt2008epje, Takano2023}. Particularly, the Leslie thermomechanical power $\text{LTP} \sim +4 \times 10^8\ \unit{\newton.\metre^{-1}.\kelvin^{-1}.{\%\, w/w}^{-1}}$ coincides with that of 7CB(\textit{R}) measured in \ChLC~phase\cite{Oswald2014epl}, which strongly supports that the rigid-body rotation in glycerol is attributed to the thermal Leslie effect.

\subsection{Model B: AZ effect under Oswald’s approximation}\label{sec:model_b}
    The reversal was analysed with the AZ effect.

\subsubsection{Basics of the Akopyan–Zel’dovich effect.~~}\label{sec:basics_of_az}
    The AZ effect is another mechanism that possibly explain reversal behaviour. The AZ effect is a cross-coupling phenomenon in which a gradient of orientation field breaks local symmetry to couple temperature gradient $\vec{\nabla} T$ with stress. Here, we revisit the AZ effect and attempt to account for the radius-dependent reversal.

    The AZ effect has traditionally been formulated in terms of stresses arising from gradients of the director field $\vec{\nabla} \vec{n}$ and $\vec{\nabla} T$. While symmetric and antisymmetric components are admissible, the antisymmetric stress is of primary significance in thermally driven rotation, where the torque is concerned. The antisymmetric stress associated with the AZ effect has been characterised by four independent coupling coefficients\cite{Oswald2017LiqCryst}. Several equivalent formulations have been proposed, differing only in the manner in which the director gradient is decomposed\cite{Akopyan1984, Oswald2017LiqCryst, Dequidt2016SoftMatter}.
    
\begin{figure}[bthp]
\centering
    \includegraphics[width=8.3cm, keepaspectratio]{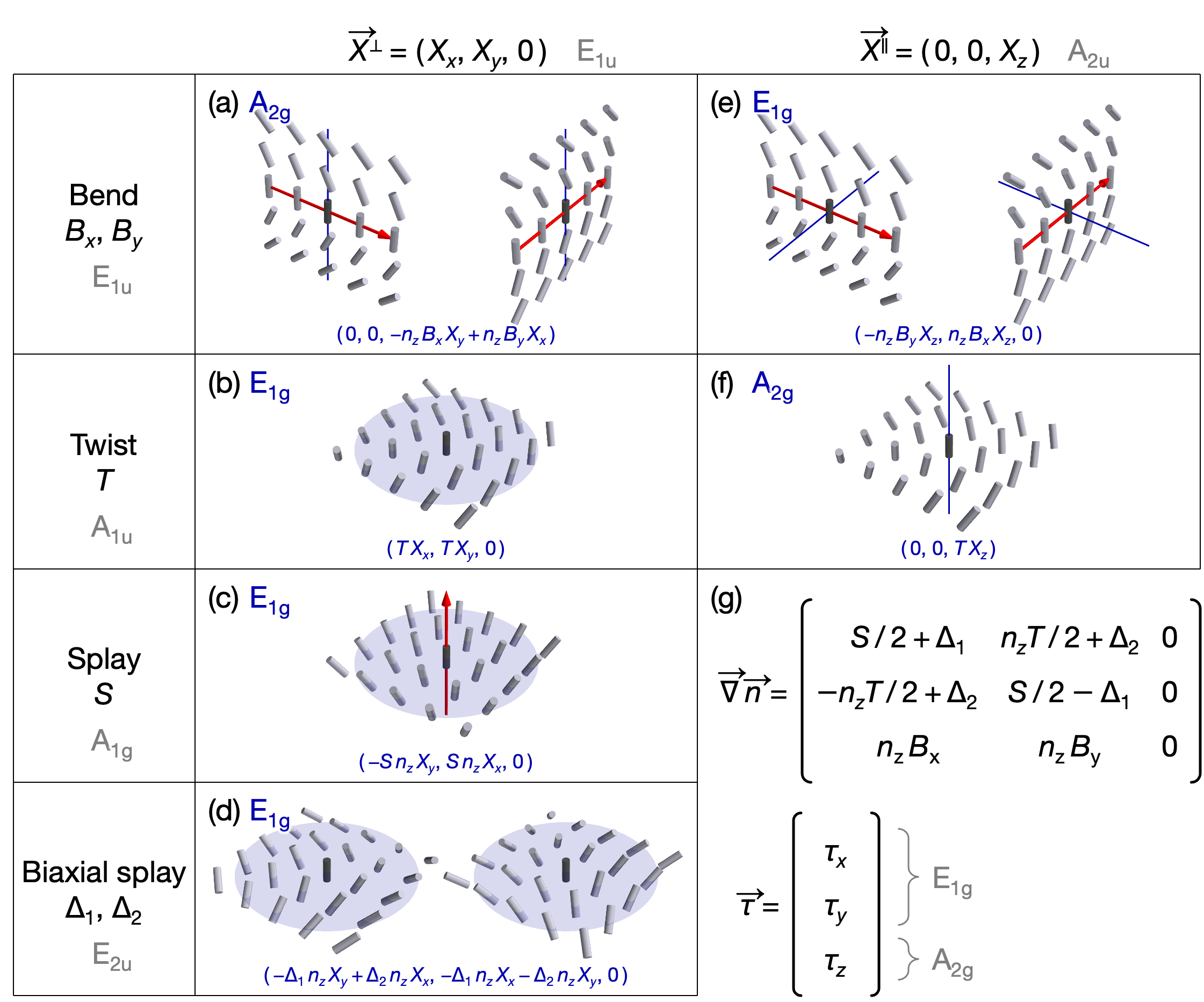}
    \caption{
        Schematics of the four ($6$~DOF) director deformation modes (See also Movies~S9--S14\cite{seeZenodo}) and six modes ($9$~DOF) of AZ torque induced at the origin, where is designated by the black cylinder. Grey cylinders represent the deformed director field around the origin. The torques lie along the blue lines or within the disc. $\vec{X}$ denotes the affinity, whereas $\vec{X}^\perp$ and $\vec{X}^\parallel$ indicate its components perpendicular and parallel to the director, respectively. 
        (a)--(f) Independent modes of AZ torque and each their symmetry is displayed on the right of each label. The blue formulae represent the torque components.
        Red arrows in panels (a) and (e) represent the bend vector $-(\vec{n}\cdot\vec{\nabla})\vec{n}$. The red arrow in panel (c) represents $\vec{n}(\vec{\nabla}\cdot\vec{n})$.
        (g) Components of the deformation $\vec{\nabla} \vec{n}$ and the torque density $\vec{\tau}$.
    }
    \label{fig:AZ_torque_modes}
\end{figure}

    We determined the independent degrees of freedom~(DOF). The gradient of the director $\vec{\nabla} \vec{n}$ is decomposed into the bases of irreducible representations: bend~($2$~DOF), splay~($1$~DOF), twist~($1$~DOF), and biaxial splay~($2$~DOF)\cite{Machon2016PhysRevX, Selinger2018, Selinger2024introduction}. The temperature gradient $\vec{\nabla} T$ is decomposed into irreducible components: $1$~DOF parallel and $2$~DOF perpendicular to $\vec{n}$. Proper combinations of these irreducible representations form the independent DOF of the torque (Fig.~\ref{fig:AZ_torque_modes}). The torque is expressed by the bases of irreducible representations of  $D_{\infty\text{h}}$. In particular, the component of the torque parallel to the director is the one-dimensional \rep{A}{2g} representation, while the perpendicular components are the two-dimensional \rep{E}{1g} representation since the torque is an axial vector. So, the torque can be decomposed into six modes (Fig.~\ref{fig:AZ_torque_modes}). Four of them in Figs.~\ref{fig:AZ_torque_modes}b--e are previously known\cite{Akopyan1984,Oswald2017LiqCryst}, and all belong to the \rep{E}{1g} representation. Our notation is converted to that of the previously reported works\cite{Akopyan1984,Oswald2017LiqCryst,Dequidt2016SoftMatter} through a linear transformation. We have newly derived two additional coefficients in the \rep{A}{2g} representation, which do not generate a molecular field but induce antisymmetric stress (Figs.~\ref{fig:AZ_torque_modes}a and f). The corresponding stress have been mentioned\cite{Brand1987PhysRevA}.

\subsubsection{Details of analysis with Model B.~~}\label{sec:detail_model_b}
    The AZ torque is characterised by six coupling coefficients. For simplicity, Oswald\etal~approximated the coefficients of the \rep{E}{1g} representation to be same in his notation and neglected the \rep{A}{2g} representation. We adopt the same approximation here. 
    
    We consider the orientational order in Model~A and the AZ effect, and this assumption is referred to as Model~B. Figure~\ref{fig:long}b shows the $R^\ast$ predicted by Model~B. In the cylinder, region between skyrmion has the opposite sign of twist. The AZ effect reflects the opposite twist and radius-dependent reversal occurs. At $R^\ast$, torque due to the AZ effect is cancelled out ($T = 0$) and we obtain
    \begin{equation}
        R^\ast
        =
        \sqrt{\frac{9}{4\pi} \int_{D_a} dxdy\ \left( 1 - \frac{p_0}{2\pi} \left[ \vec{n}^1 \times \nabla_z \vec{n}^1 \right]_z \right)}
        ,
    \end{equation}
    where $D_a$ denotes the circular base of the cylinder.

    In $\delta = 0.25$, $R^\ast$ did not go through the grey region and reversal is impossible (Fig.~\ref{fig:sixpanels}b). Reversal is possible in greater distance like $\delta = 0.5$. Greater $\delta$ brings to greater volume of region with opposite twist which causes the reversal. Note that, $R^\ast$ is parameterised by $a$ and $\delta$ but cross-coupling coefficient $\bar{\xi}$ never affects $R^\ast$.
    
    Figure~\ref{fig:long}d shows experimentally obtained $R^\ast / p_0$ against $T - T^\ast$. The reversal possesses perfect mirror symmetry as $R^\ast / p_0$ coincides among 7CB(\textit{R}) and 7CB(\textit{S}). $R^\ast / p_0$ varied $1\text{--}2$ in temperature-dependent manner. Model~B reproduced $R^\ast$ whose order of magnitude was consistent with the experiments when $\delta = 0.5$. For better prediction, several modifications will be possible. The dependence of the coupling coefficients on $S$ should be incorporated. The scalar OP is likely lower around the disclination than in the shell, and this variation should be reflected in the AZ effect. By reformulating the AZ effect using the $Q$-tensor formalism, the dependence on $S$ is naturally included\footnotemark[2]. All six modes of the AZ effect should be considered since their coefficients may depend on $S$ in a different manner. Though Oswald's approximation assumes that \rep{A}{2g} modes (Figs.~\ref{fig:AZ_torque_modes}a and f) were absent and the other modes have the same sign, symmetry allows all modes to occur and have different signs.
    
    Even with modifications, the AZ effect cannot account for the reversal. The AZ effect converts thermal energy into mechanical work. Since its conversion efficiency cannot exceed the Carnot limit, the coupling coefficients are subjected to constraints. The induced stress is proportional to the irreducible modes of $\vec{\nabla} \vec{n}$. Since the deformation is not intrinsic to the material and can be arbitrarily large, the induced stress may increase infinitely as the deformation increases. This situation violates the Carnot limit. Thus, the coupling coefficients must be equal to zero: the AZ effect is forbidden by thermodynamics\footnotemark[2]. In contrast, the flexoelectric effect\cite{Meyer1969PhysRevLett, Dequidt2007EurPhysJE} is possible, as it is a reversible phenomenon. Although some previous reports have claimed experimental observation of the AZ effect\cite{Ignes-Mullol2016prl, Oswald2017LiqCryst}, these experiments were conducted in phase-separated systems of miscible lyotropic mesogen and solvent. The observed rotation should be reanalysed as an optical illusion caused by Marangoni convection\cite{Oswald2019softmatter} due to a large interfacial entropy. While the AZ effect is an attractive hypothesis from a symmetry perspective, it does not constitute a consistent mechanism.

\subsection{Model C: modified Leslie effect}\label{sec:model_c}
    Models~A and B have no ability to account for the temperature dependence of the reversal. This failure originates from that they do not incorporate the temperature dependence of the cross-coupling coefficients. As is the case for the elastic constants\cite{de_Gennes}, thermal conductivity\cite{Marinelli1998PRE}, and viscosity\cite{Cui1999MolCrystLiqCryst}, the coupling coefficients are expected to depend on the scalar OP $S$, and hence on temperature. By applying the $Q$-tensor formalism, it has been shown that the Leslie effects can incorporate the $S$ dependence of the cross-coupling coefficients\cite{Takano2025PhysRevE}. This theory predicts that the Leslie effects vanish in the \ILC~phase and also indicates that a heat current parallel to $\vec{n}$ induces a torque. Since angular momentum and heat current are ``true-chiral'' objects in Barron's terminology\cite{Barron2012chirality}, such an $\vec{n}$-independent Leslie effect is natural. Indeed, the torque component parallel to $\vec{n}$ belongs to the \rep{A}{2g} representation, and is therefore symmetry allows it. Furthermore, this effect is consistent with the second law of thermodynamics\cite{Takano2025PhysRevE}

    Both traditional $\vec{n}$-dependent and new -independent thermal Leslie effects as well as their $S$ dependence were considered. The torque density is given by
    \begin{equation}
    \label{eq:tau_nu2_S2_nu_3_S3_nu3_S3}
        \vec{\tau}
        =
        \left( \nu_2 S^2 + \nu_3 S^3 \right) \vec{n} \times ( \vec{n} \times \vec{\nabla} T )
        + \nu_3' S^3 \vec{\nabla} T
        ,
    \end{equation}
    here the lowest exponent of $S$ is determined by \OnsagerVariational\footnotemark[2]. We adopted the director field in Model~A and obtained
    \begin{subequations}
    \label{eq:model_c}
    \begin{align}
        R^\ast
        &=
        \sqrt{\frac{9}{8\pi} g(S) \int_{D_a} dxdy\ \left( {n^1}_z \right)^2}
        ,\\
        g(S)
        &=
        \frac{1 + \frac{\nu_3}{\nu_2}S}{1 + \frac{\nu_3}{\nu_2} \left(1 + \frac{3}{2} \frac{\nu_3'}{\nu_3}\right) S}
        .
    \end{align}
    \end{subequations}
    Figure~\ref{fig:sixpanels}c shows $R^\ast$ against $a$. Significant $R^\ast$ is possible for a proper $g$ under feasible parameters $\delta = 0.25$ and $a = 0.5$, whose values are suggested by simulation\cite{Sec2012}.  The factor $g$ varies with $T$ through an explicit $S$ dependence. Since the power law holds for the $T$ dependence of $S$\cite{Haller1975}, we assumed $S \propto
    \left( T^\ast - T + \delta T\right)^\beta$, where the coefficients $\delta T \sim 1\ \unit{\degreeCelsius}$, $\beta \sim 0.2$ were chosen by considering the linear birefringence of 5CB\footnotemark[2]. The positive $\delta T$ indicates that the \ChLC--\ILC~transition is of the first order. The experimentally obtained $R^\ast$ was fitted by Eq.~\eqref{eq:model_c} (Fig.~\ref{fig:long}d). Model~C has successfully reproduced the increase in $R^\ast$ with rising temperature. The reversal and its $T$ dependence requires both $S$-dependent and $\vec{n}$-independent thermal Leslie effect. The fitting parameters $S \nu_3 / \nu_2$ and $\nu'_3 / \nu_3$ enable us estimated the Leslie coefficients in Eq.~\eqref{eq:tau_nu2_S2_nu_3_S3_nu3_S3}. We have already estimated $\nu$ for singular RSS at low temperature $T^\ast - T + \delta T \sim 10\ \unit{\kelvin}$ where molecules are supposed to be ordered well: $S \sim 0.6$. Under these assumption, we estimated as
    \begin{subequations}
    \begin{align}
        \nu_2
        &=
        \left\{
            S^2
            \left(
                1 + \frac{S \nu_3}{\nu_2} 
                \left[
                    1 + \frac{3}{2} \frac{\nu'_3}{\nu_3}
                \right]
            \right)
        \right\}^{-1}
        \nu
        \sim
        - 5 \times 10^{-6}\ \unit{\newton\metre^{-1}\kelvin^{-1}}
        ,\\
        \nu_3
        &=
        \frac{S\nu_3}{\nu_2}S^{-1}\nu_2
        \sim
        +3 \times 10^{-5}\ \unit{\newton\metre^{-1}\kelvin^{-1}}
        ,\\
        \nu'_3
        &=
        \frac{\nu'_3}{\nu_3} \nu_3
        \sim
        - 1 \times 10^{-5}\ \unit{\newton\metre^{-1}\kelvin^{-1}}
        .
    \end{align}
    \end{subequations}
    The present analysis indirectly estimated those value in the first time. The new coefficient $\nu'_3$ is comparable to the known coefficient $\nu$ and sufficiently large to be measured. Future studies would address to obtain $\nu'_3$ by direct experiments. 
    
    Further precise reproduction of $R^\ast$ would be achieved by considering the actual director configuration. To compliment the differences in the orientational field between the disclination and the shells, it is necessary to incorporate the more precise fractional contributions of each irreducible mode.

\section{Conclusions}
    We have demonstrated that the direction of the thermally driven rotation in cholesteric liquid crystal droplets reverses depending on droplet size and temperature. This reversal cannot be attributed to the handedness of the cholesteric helix or explained by the conventional description of the thermal Leslie effect. Instead, the combined experimental and phenomenological analyses consistently indicate that the Leslie thermomechanical coupling coefficient may be possibly influenced by the scalar order parameter and may even change its sign as the orientational order varies. These findings establish an experimentally accessible route to manipulating thermomechanical coupling through the liquid crystal state itself.
    
    Beyond identifying a previously unrecognised reversal mode, the present work provides a broader physical perspective on thermomechanical energy conversion in condensed matter. Thermomechanical coupling has generally been regarded as an intrinsic material property determined by molecular chirality and mesoscopic structure. Our results suggest that it should instead be viewed as an emergent property of the non-equilibrium ordered state, which can be continuously tuned through changes in molecular ordering. The scalar order parameter therefore represents not merely a descriptor of the nematic order, but an active design parameter governing the direction and magnitude of thermomechanical energy conversion. This perspective extends the current understanding of the thermal Leslie effect and offers a general framework for designing responsive chiral soft materials with programmable thermomechanical functionality.
    
    Our results suggest that thermomechanical coupling should not be regarded as a fixed material characteristic, but rather as a state-dependent property emerging from molecular ordering under non-equilibrium conditions. The ability to convert small temperature gradients directly into controlled rotational motion provides a fundamental design principle for exploiting ubiquitous low-grade thermal energy. Rather than relying on multistep energy-conversion processes, thermomechanical coupling enables heat to be transformed directly into organised mechanical motion through molecular ordering. Such a mechanism offers promising opportunities for passive micro-actuation, autonomous transport and adaptive microsystems operating under ambient thermal gradients. More generally, our findings highlight how controlling the orientational order can expand the functionality of soft materials, providing new strategies for sustainable energy utilisation through direct heat-to-motion conversion.

\section{Experimental}
    The research methodology is explained in the main text, and further information is available in the SI\footnotemark[2].

\section*{Author contributions}
    S.T.: conceptualisation; formal analysis; funding acquisition; investigation; methodology; software; visualisation; writing -- original draft; writing -- review \& editing,
    T.N.: conceptualisation; supervision; writing -- review \& editing,
    K.N.: project administration; writing -- review \& editing,
    T.A.: Conceptualisation; resources; supervision; writing -- review \& editing

\section*{Conflicts of interest}
    There are no conflicts to declare.

\section*{Data availability}
    The data analysis scripts and generated data, including the supplementary movies, are available on Zenodo via the following DOI:~10.5281/zenodo.22112415. The experimental and computational methodology, sample composition, modelling derivation, and other supporting information for this article are provided in the Supplementary Information (SI)\footnotemark[2]. References
\cite{de_Gennes1972,Gray1975,Takano2023,Yoshioka2018,Oswald2019softmatter,Bouligand1984,Sec2012,Rika2012,Pestov2018,Poy2019,Meng2023,Haller1975,Li2005,Yamaguchi1988,seeZenodo,Lamb1895,Yamamura2015EKISHO,Takano2025PhysRevE,de_Gennes,Brand2018RheolActa,Pleiner1996,Onsager1931I,Onsager1931II,Parodi1970,Kubo1957,Tovkach2017,Akopyan1984,Brand1987PhysRevA,Oswald2017LiqCryst,Dequidt2016SoftMatter,Poy2018,Doi2011,Barron2012chirality,Barato2015,Macieszczak2018prl,Leslie1979,KlemanLavrentovich2003,Buckingham1914,Cheng2008IndEngChemRes,Machon2016PhysRevX,Selinger2018,Bezic1992} are cited in the SI. Other data are available from the corresponding author upon reasonable request.

\section*{Acknowledgements}
    S.T. deeply thanks Y.~Tabe for providing the reagents and instrumentation, Y.~Maruyama for helpful comments at the early stage, Y.~Yamamura for insightful discussions on the conformations of $n$CB molecules, W.~Mori and S.~Aoki for practical suggestions to optimise the simulation code, and N.~Mukai for confirming the bimeron structure.
    This work was supported by JST~SPRING, Grant No.~JPMJSP2128; JSPS Grant-in-Aid for JSPS Fellows, Grant No.~JP25KJ2180; and Waseda Research Institute for Science and Engineering Grant-in-Aid for Young Scientists (Early~Bird) to S.T. The travel expenses for S.T.'s stay during the WPI-SKCM$^2$ InternKNOTship Program were funded by WPI-SKCM$^2$.



\balance


\bibliography{clan,si_repository} 
\bibliographystyle{rsc} 

\end{document}